\documentclass[%
reprint,floatfix,
 amsmath,amssymb,
 aps,
pra,
]{revtex4-2}

\usepackage{graphicx}
\usepackage{dcolumn}
\usepackage{bm}
\usepackage{multirow}
\usepackage{caption}
\usepackage{blindtext}
\usepackage[dvipsnames]{xcolor}
\usepackage{graphicx}
\usepackage{subfig}
\begin{document}

\title{A novel method for measuring the Fermi velocity of elemental targets}
\author{Manpreet Kaur$^1$ and T. Nandi$^{2*}$}
\affiliation{$^1$Department of Physics, University Institute of Sciences, Chandigarh University, Gharuan, Mohali, Punjab 140413, India}
\affiliation{$^{2}$Department of Physics, Ramakrishna Mission Vivekananda Educational and Research Institute, PO Belur Math, Dist Howrah 711202, West Bengal, India}
\thanks {Email:\hspace{0.0cm} nanditapan@gmail.com (corresponding author). Past address: Inter-University Accelerator Centre, JNU New Campus, Aruna Asaf Ali Marg, New Delhi-110067, India.}
\begin{abstract}
The right kind of theoretical treatment of direct Coulomb ionization of inner-shell of target atoms including multiple ionization of their outer-shells by using accurate x-ray fluorescence yield data and electron capture by projectile ions from inner-shell electrons of target atoms enables us to fully understand the complex physics issues with the heavy-ion-induced inner-shell ionization phenomenon. Such great success has only been achieved recently [Phys. Rev. A 111 (2025) 042827]. Aftermath, further investigations exhibit such a picture only if the Fermi velocity of the elemental target is accurate, as it takes a significant role in correct evaluation of charge-state distribution of the projectile ions inside the target, which contributes an invaluable share in calculating the electron capture-induced ionization cross section correctly. In this work,  we devise a powerful method that enables us to measure the correct and accurate Fermi velocity for almost every elemental metal in the periodic table. As per our present knowledge, this in turn not only improves our understanding of the said complex physics issues one step ahead but also helps move toward further miniaturization of integrated circuits and use the heavy-ion-induced X-ray emission in impurity analysis more reliable and accurate.
\end{abstract}

\maketitle

\section{Introduction}

The further miniaturization of integrated circuits (IC) \cite{sreenivasan2017nanoimprint} is a significant field of current research and developments. This can be possible by using the narrower interconnect lines. However, the electrical resistivity of metal wires increases with decreasing width. This fact represents a major challenge for further down-scaling of integrated circuits. Normally, Cu interconnects are used in IC and the search for other metals that have lower resistivity is the focus of theoretical calculations. However, one cannot calculate the electrical resistivity of the metallic interconnects; nevertheless, one can study this property through the products of $\lambda\rho_0$ (=$\frac{mv_F}{ne^2}$) and $\tau\rho_0$ (=$\frac{m}{ne^2}$), where $\lambda$ is the mean free path of the electron, $n$ is the electron density, $v_F$ is the Fermi velocity, $e$ is the electronic charge, $m$ is the mass of an electron, $\tau$ the average relaxation time, and $\rho_0$ is the resistivity at room temperature. In addition to these products, the average Fermi velocity $v_F=\frac{\lambda}{\tau}$ can also be used. Free electron gas (FEG) model \cite{Ashcroft76} is often used to calculate $v_F$. Besides, \citet{gall2016electron} has calculated $v_F$ for the 20 most conductive elemental metals by numerical integration on the Fermi surface obtained from first principles, using constant values of the approximations for $\lambda$ and $\tau$ and the wave vector ($k$) dependent $v_F (k)$. Subsequently, \citet{montanari2017low} used a non-perturbative approximation to the electronic stopping power based on the central screened potential of a projectile moving in a free-electron gas to calculate $v_F$ for ten conductive metals. Hence, it is a modified FEG model. We notice that the results of \citet{gall2016electron} differ from both FEG \cite{Ashcroft76} and modified FEG \cite{montanari2017low} models for the two metals Be and Al for which both the refs.\cite{gall2016electron,montanari2017low} have calculated and shown in Table \ref{Tab:Gall-Montanari}. Here, we can see that the FEG model predictions differ very much from Gall model for both the metals. Whereas the modified FEG model \cite{montanari2017low} prediction for Be is almost equal to FEG model calculation, but this is not so for Al. It is highly imperative to verify these predictions \cite{gall2016electron,montanari2017low,Ashcroft76} through different metals by a precision experiment so that search for higher conductivity of interconnect lines of the metals than that of copper can proceed in the right directions.
\begin{table}[htbp]
    \centering
    \caption{Comparison of Fermi velocities $v_F$ for two metals (Be and Al) with two recent theoretical studies \cite{gall2016electron,montanari2017low} and the electron gas model \cite{Ashcroft76} also. The values for $v_F$ are given in 10$^6$ m/s units. }
    \label{Tab:Gall-Montanari}
\begin{tabular}{lccc}
    \hline
    El. & {\hskip 0.8cm}Ref.\cite{gall2016electron}{\hskip 0.8cm} & Ref.\cite{montanari2017low} & {\hskip 0.8cm} Ref.\cite{Ashcroft76}\\
    \hline
    Be & 1.262  & 2.245 &{\hskip 0.8cm}2.25\\
    Al &  1.599 &  2.921&{\hskip 0.8cm} 2.03\\
    \hline
\end{tabular}
\end{table}
\par
We propose here a new precision measurement method, which can be applicable to any elemental metal in the periodic table, regardless of whether its Fermi surface is spherical, non-spherical, or anisotropic. In this paper, we first describe the measurement method and then demonstrate this method to measure $v_F$ in several elemental metals. Finally, we compare the experimental results presented here with available experimental and theoretical results to discuss current status and future challenges.

\begin{figure}
    \centering
    \includegraphics[width=8.0cm,height=10.0cm]{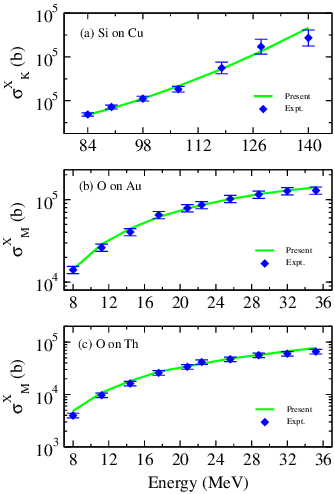}
    \caption{Comparison of experiment \cite{chatterjee2021significance, CZARNOTA2009-PhysRevA.79.032710} and theory \cite{kaur2023understanding} for K or M x-ray production cross section as a function of beam energies: For K x-ray production cross section of copper by silicon impact (a), M x-ray production cross section of gold by oxygen impact (b) and M x-ray production cross section of thorium by oxygen impact (c). Used $v_F$ values in the calculation were 1.110$\times 10^6$ m/s \cite{gall2016electron}, 1.382 $\times 10^6$ m/s \cite{gall2016electron} and 1.402 $\times 10^6$ m/s \cite{Ashcroft76} for copper, gold and thorium, respectively.}
    \label{Fig: Cu-Au-Th}
\end{figure}
\begin{figure}
    \centering
    \includegraphics[width=8.5cm,height=10.0cm]{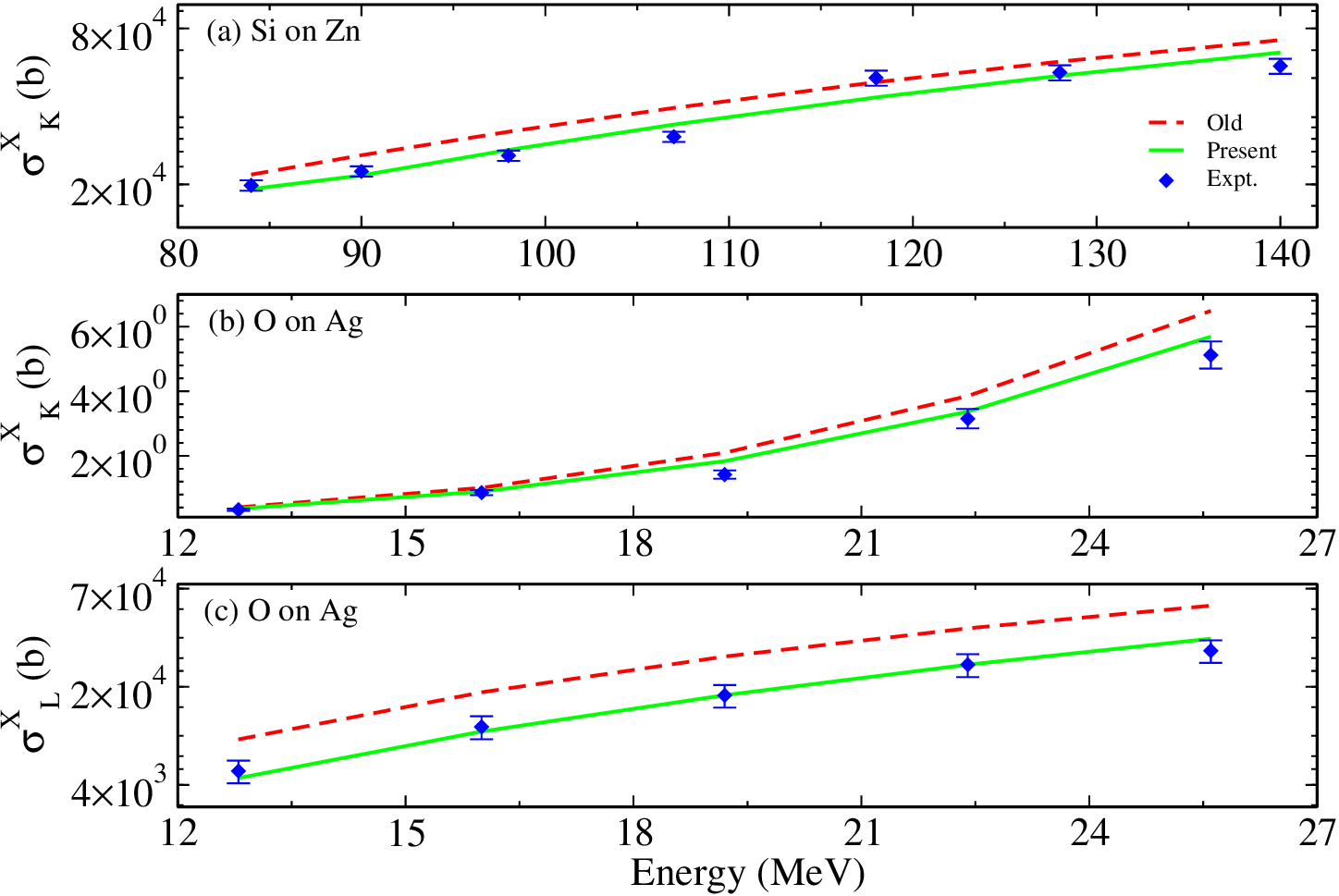}
    \caption{Measurement of Fermi velocity from x-ray production cross section as a function of beam energies: (a) Comparison of experimental K x-ray production cross section of zinc target atoms by silicon projectiles \textcolor{red}{\cite{chatterjee2021significance}} is made with theoretical estimates using the formalism of \citet{kaur2023understanding}. The theoretical curve with the red dashed line using the $v_F$ value of 1.566$\times10^6$ m/s \cite{gall2016electron} differs considerably from the experimental data. Good alignment of the experiment with theory was achieved by varying the $v_F$ value. Final iteration is shown with solid green line and thus the measured $v_F$ is 1.735 $\times 10^6$ m/s. (b) Similarly, experimental K x-ray production cross section of silver target atoms by oxygen projectiles is compared with theory using the $v_F$ value of 1.440 $\times10^6$ m/s \cite{gall2016electron}. The difference between experiment \textcolor{red}{\cite{GORLACHEV201634}} and theory is resolved by tuning $v_F$ and final $v_F$ is found to be 3.710 $\pm$ 0.090 $\times 10^6$ m/s. (c) Comparison between experiment and theory is made with L x-ray production cross section of silver target atoms by oxygen projectiles. The observed difference is sorted out with the $v_F$ value determined with the K x-ray production cross sections as shown in Fig. (b).  }
    \label{fig:zn-ta}
\end{figure}
\begin{figure}
\centering
\includegraphics[width=8.0cm,height=7.0cm]{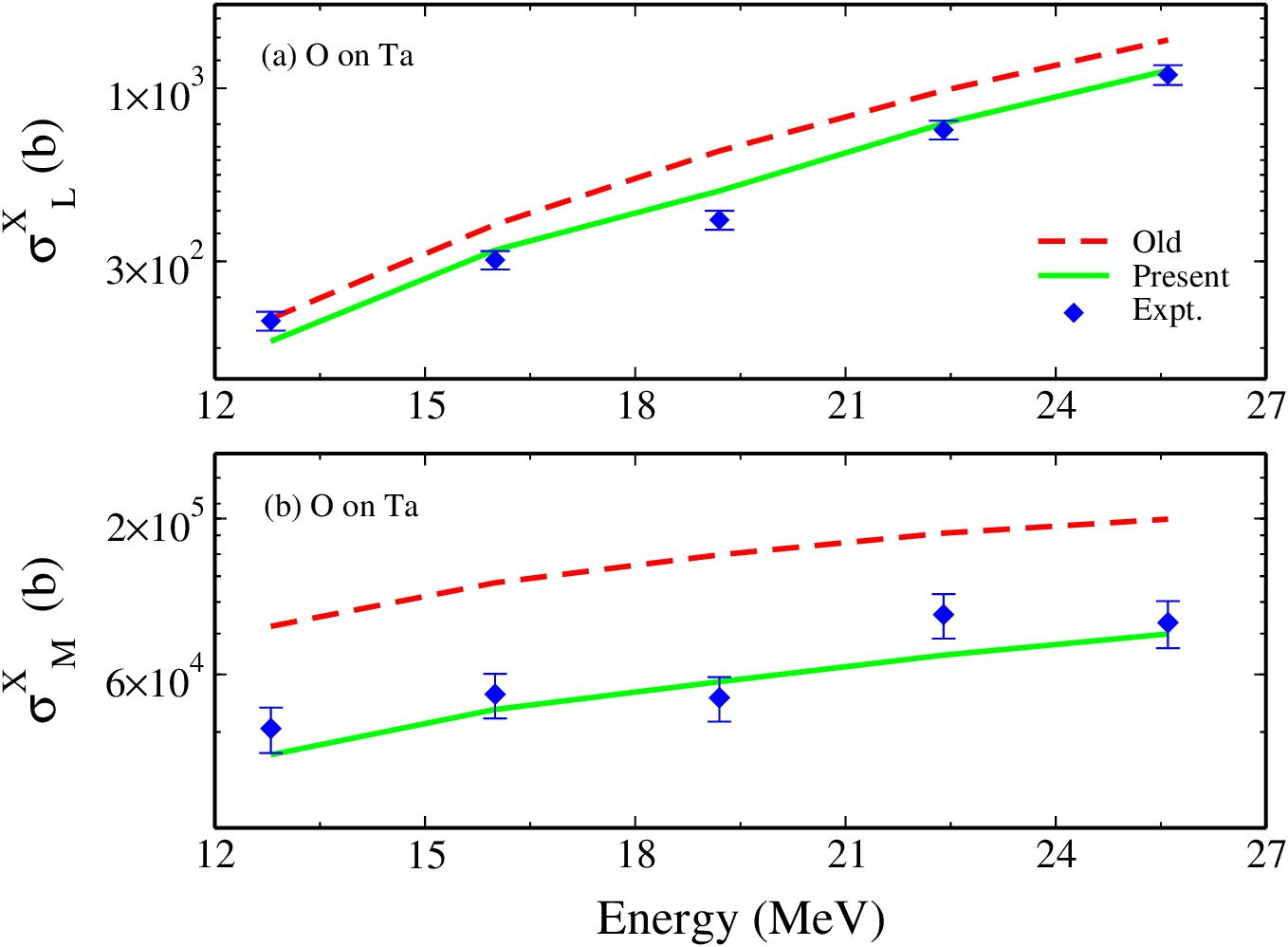}
\includegraphics[width=8.0cm,height=7.0cm]{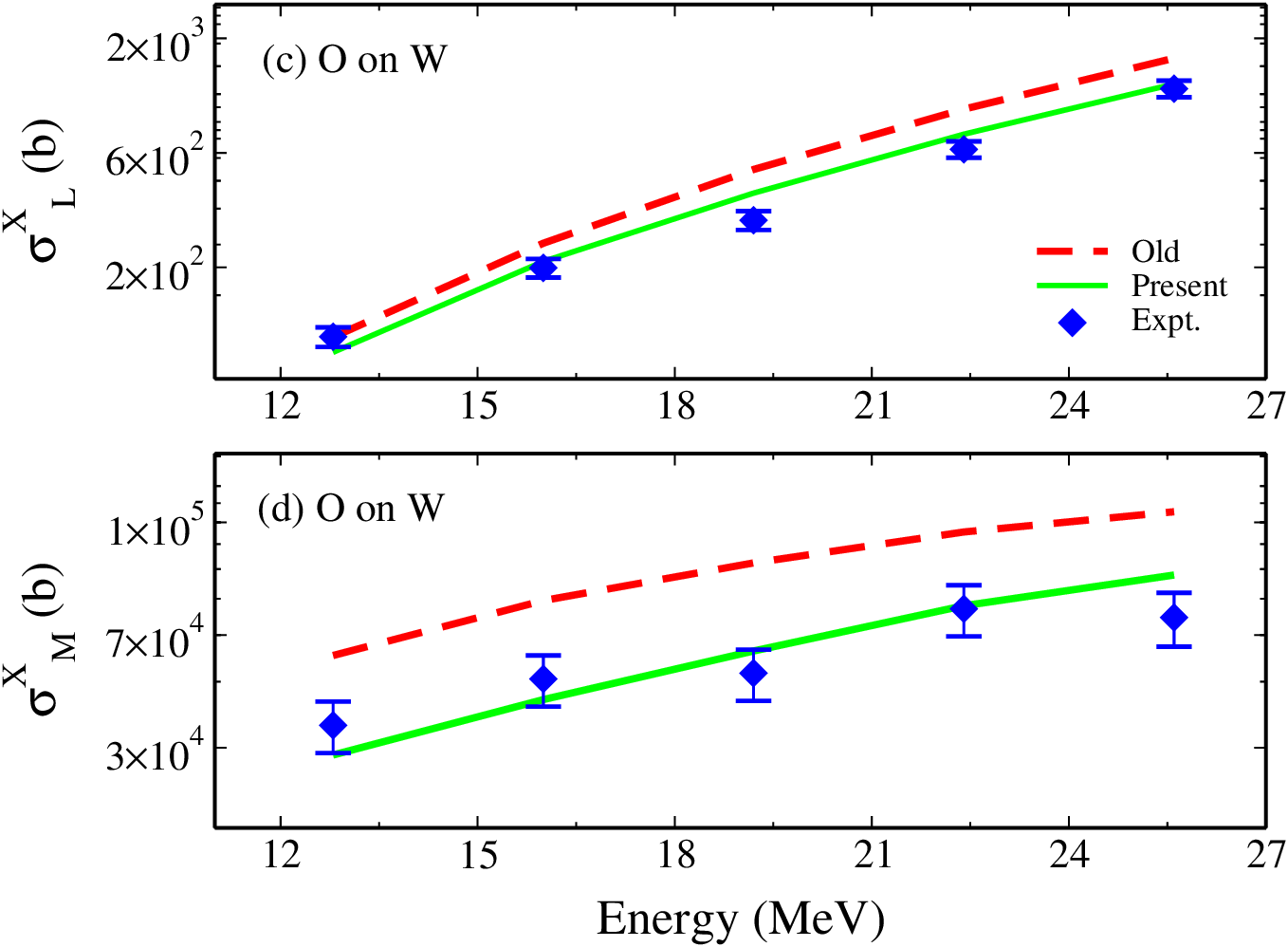}

\caption{Measurement of Fermi velocity from x-ray production cross section as a function of beam energies for two more targets, tantalum and tungsten \textcolor{red}{\cite{GORLACHEV201634}}: Fermi velocity of tantalum is determined using L and M x-ray production cross sections in Fig. (a) and (b), respectively. Difference between experiment and theory observed with $v_F$=2.340 $\times 10^6$ m/s \cite{gall2016electron} is resolved with $v_F$=3.488 $ \pm 0.056 \times 10^6$ m/s. Next, Fermi velocity of tungsten is determined using L and M x-ray production cross sections in Fig. (c) and (d), respectively. Difference between experiment and theory observed with $v_F$=0.971 $\times 10^6$ m/s \cite{gall2016electron} is resolved with $v_F$=3.924 $ \pm 0.064 \times 10^6$ m/s. } 

\label{fig:Ag-W}
\end{figure}

\section{Proposed experimental method}
Complex problems of inner-shell ionization of heavy target atoms by energetic heavy ions have remained theoretically unsolved for decades due to (i) multiple ionization (MI) in the outer shells of target atoms \cite{lapicki1986multiple,banas2002multiple,lapicki2004effects,msimanga2016k}, (ii) inner-shell vacancy creation in target atoms due to electron capture (EC) by projectile ions \cite{lapicki1977electron,lapicki1980electron,Meyerhof} in addition to direct Coulomb ionization (DCI) \cite{basbas1973universal,lapicki1979coulomb, lapicki1980coulomb,brandt1981energy,lapicki1989cross} and (iii) incorrect fluorescence yields \cite{singh2002m,jones2007x, ager2017reconsidering}. In the last couple of years, all these problems have been solved one after another to great extent by our group \cite{chatterjee2021significance,chatterjee2022understanding,kaur2023understanding, kaur2025towards}.  We did not stop here; rather we kept on studying many more cases to verify the said success story in length and breadth. We encounter certain shortcomings with the estimations of the contribution of the electron capture-induced ionization part.
\noindent
To evaluate the inner-shell ionization cross section due to electron capture by projectile ions, one can use the Oppenheimer-Brinkman-Kramers (OBK) approach as described by \citet{lapicki1980electron}. A detailed description can be seen in \cite{kaur2023understanding}. One significant point to note is that inner-shell ionization due to DCI-MI or EC takes place inside the target foil. Moreover, the charge state fractions ($F(q)$) of the projectile ions which are required to evaluate the EC induced ionization cross section must be within the target and it is imperative to know them very well. This can only be possible if we can get the mean charge state inside the target foil ($q^i_m$) very accurately.
To obtain the $q^i_m$, we employ the Fermi gas model (FGM) \cite{brandt1973dynamic}  as follows: 
\begin{equation}
q_m^i = Z_1(1-\frac{v_F}{v_1})
\label{q_m^i}
\end{equation}
where $ Z_1$ and $v_F$ are the projectile atomic number and Fermi velocity of target electrons, respectively. This formula works only well if the projectile ion velocity ($v_1$) is greater than $v_F$, and the maximum stripping occurs when $v_1 >> v_F$. Furthermore, this formula is applicable only for the foil thickness $\ge$ the equilibrium target foil thickness. Note that the accuracy of estimated $q_m^i$ depends solely on the accuracy of the values of $v_F$.
\par
Now, accurate $q_m^i$ and the right choice of charge-state distribution \cite{sharma2016x} and the corresponding distribution width led us to accurately calculate the ionization cross section due to electron capture as described in \citet{kaur2023understanding}. Such ionization cross sections can now be converted to the x-ray production cross section using accurate values of fluorescence yields \cite{kaur2025towards}. A sum of the x-ray production cross section due to DCI-MI and EC gives us a total theoretical x-ray production cross section, which can be compared to the total experimental x-ray production cross section. Our detailed study showed that an excellent agreement between theory and experiment is achieved if $q_m^i$ or more specifically $v_F$ is accurate. This happens irrespective of whether the x-ray production cross section is considered for the K-, L-, or M-shell.  Hence, comparing the experimentally measured x-ray production cross section as a function of projectile energy with the theoretical counter part using the preciously known $v_F$ values can verify the proposed method. This exercise has been used through the case of two noble metals, copper and gold and one case of actinide metals, thorium, as shown in Fig. \ref{Fig: Cu-Au-Th}. 
\par
Having seen an excellent match, we proceed to devise a Fermi velocity measurement  method using the experimentally obtained heavy-ion induced x-ray production cross sections as a function of beam energy as follows: (i) If we find that the experimental x-ray production cross section curve differs from the corresponding theoretical curve, (ii) we ought to vary the $v_F$-value that is used in theoretical calculations till we achieve an excellent agreement between experiment and theory, (iii) Final value of this $v_F$ represent the measured $v_F$, (iv) If this measured $v_F$ fits well with another x-ray production cross section data set, then the measurement is said to be validated experimentally, and (v) Corresponding measurement uncertainty is related to the measurement uncertainty of the XPCS measurements. 
\par
The proposed method has been demonstrated from the XPCS data of (i) zinc by the impact of silicon ions and (ii) silver by the impact of oxygen ions shown in Fig. \ref{fig:zn-ta}. The theoretical calculation shown with the red dashed line using the value of $v_F$ of 1.566$\times10^6$ m/s \cite{gall2016electron} differs considerably from the experimental data (Fig. \ref{fig:zn-ta}(a)). Good alignment of the experimental data with the theoretical predictions was achieved by varying the $v_F$ value in the calculation shown with the green line and the measured $v_F$ turns out to be 1.735$\pm$ 0.090 $\times 10^6$ m/s. A similar exercise was also followed for the silver case. Here we use K and L x-ray production cross section data as a function of beam energy and the measured $v_F$ is found to be 3.710$\pm$ 0.090 $\times 10^6$ m/s.
\par
Merit of above mentioned method can be validated to a higher degree through the measurements when the x-ray production cross section data include more than one type of characteristic x-rays. For example, K and L or L and M or even K, L and M x-ray production cross-section measurements are done for a particular projectile-target system. Three of such cases are shown in this paper. Here, both the K and L x-ray production cross-section measurements are shown for silver  in Fig.\ref{fig:zn-ta} and both the L and M x-ray production cross-section measurements for tantalum and tungsten are shown in Fig.\ref{fig:Ag-W}. All these experiments were carried out by oxygen ion impacts. Interestingly, $v_F$ obtained from the K x-ray production cross-section data fits well with the L x-ray production cross-section data for silver. Similarly, $v_F$ obtained from the L x-ray production cross-section data aligns well with the M x-ray production cross-section data of tantalum and tungsten.
\section{Results and discussions}
Although the above-mentioned measurement method is potentially applicable to any solid elemental metals, in the present demonstration, we have studied only for a small number of metals and the present experimental results obtained are compared with various theoretical estimates \cite{Ashcroft76,gall2016electron,montanari2017low} in Table \ref{tab:fermi_velocities}. Let us first describe the free-electron gas (FEG) model  \cite{Ashcroft76}. This model treats the valence electrons in a metal as a gas of non-interacting particles moving freely within the material. Thus, it simplifies the complex behavior of electrons in a metal as the independent particles, analogous to a gas. In this model, the Fermi energy ($E_F$) in $J$ is given by \cite{Ashcroft76,eisberg1985,kittel1986}:
\begin{equation}
E_F = \frac{\hbar^2}{2m_e}(3\pi^2 n)^{2/3}
\end{equation}
where $\hbar = 1.0545718\times10^{-34}$ J.s is the reduced Planck constant, $m_e = 9.10938356\times 10^{-31}$ kg is the mass of the electron and $n$ = the number density of electrons in $m^{-3}$, which is given by:
\begin{equation}
n = \frac{z \cdot \rho \cdot N_A}{A}
\end{equation}
where $z$ is the number of conduction electrons per atom, $\rho$ is the density of the material in kg/m$^3$,  $N_A$ = 6.022 $\times$ 10$^{23}$ mol$^{-1}$ is the Avogadro number, and $A$ is the molar mass in {kg/mol}. Note that $z$ used in the calculation of $v_F$ here is the same as the number of valence electrons as represented by the electronic configuration.
\par
The {Fermi velocity}  $v_F$ in m/s is given by \cite{Ashcroft76,eisberg1985,kittel1986}:
\begin{equation}
v_F = \sqrt{\frac{2E_F}{m_e}}
\end{equation}
\citet{montanari2017low} used a slightly modified free-electron gas model which is based on the velocity dependent central screened potential of a projectile moving in a free-electron gas as proposed by \citet{nagy1998scattering}. While, \citet{gall2016electron} used a different model; it assumed both electron mean path and carrier relaxation time are constant and $v_F$ is wave-vector dependent.Thus, in this model, no concept of valence or conduction electrons is used, here we have evaluated $z$ by substituting the $v_F$ obtained from this theory in FEG model to compare the variation of $z$ with each other. Similar exercise is done with the results obtained from the present experiment also and the $z$ so derived are compared in Table \ref{tab:fermi_velocities}. The comparison gives us three categories with respect to the comparison between the present measurement and FEG model predictions. In the first, the measured figures for $v_F$ are well aligned with the FEG model prediction. In the second, the measured $v_F$ values are much higher than the FEG model predictions. While, in the third category, the measured $v_F$ values are much lower than the FEG model predictions.  On obvious reasons, modified FEG model predictions \cite{montanari2017low} available for two metals Pb and Ge are close to the FEG model calculation. Out of the  eleven measurements made in this study, the Gall model \cite{gall2016electron} predictions are available for five metals. Among these five values two metals Zn and Au fall in the first, two other metals Ag and w in the second and one metal Cu in the third category.  Although Gall model prediction for Au is closed to our measurement as well as the FEG model calculation, but its prediction for Zn is about 10 and 15\% lower than our measurement and FEG model calculation, respectively. In the second category, Gall predictions for Ag is quite close to the FEG prediction, but its prediction for W is about 2.7 times lower than the FEG calculation. The scenario is even worse when we compare the Gall predictions with our measurements; the Gall predicted $V_F$ for Ag and W are about 2.6 and 4.0 times lower than the measured values. In contrast, coming to the third category the Gall prediction is extremely well aligned with our measurements for Cu. 
\par
Let us now compare the $z$ values as obtained from the model predictions and the measurements. The trend is very similar, but comparison on $z$ shown also in Table \ref{tab:fermi_velocities} is more magnified as $z$ is proportional to $v_F^{3}$. We
 can see here that $z$ value as it comes from the electronic configuration is compared with the average number of conduction electron $z^\prime$ and $z^{\prime\prime}$  as obtained from model predicted $v_F$ \cite{gall2016electron,montanari2017low} and the presently measured $v_F$, respectively. A good agreement between $z$ and $z^{\prime\prime}$ is observed for all the metals lying in the first category. $z^\prime$ differs from $z$ and $z^{\prime\prime}$ for only Zn. Coming to the second category, a drastic difference is noticed between $z$ and $z^{\prime\prime}$. $z^{\prime\prime}$ is 2 to 19 times higher than $z$. Maximum difference is found for Ag; $z$ and 
 $z^{\prime\prime}$ are in the ratio of 1:19. It implies that all the electrons in $n=4$ shell along with 5$s^1$ take part as the conduction electrons. In the third category, certain interesting features are noticed. Both $z^\prime$ and $z^{\prime\prime}$ are seen to be only 0.35 compared $z$=1 for Cu. In case of thorium, $z$ can be 2 or 4, present experimental value $z^{\prime\prime}$ is found to 2. Similar observation is noticed in case of uranium too. It can have $z$=2-6, but our experimental value $z^{\prime\prime}$ is very closed to 2 as shown in Table \ref{tab:fermi_velocities}.
 \par 
 One important point is noteworthy here that above mentioned facts on the conduction electrons cannot be attributed to any effects of heavy-ion collisions. Because the results shown were done either by silicon or oxygen impacts. All the target copper, zinc and germanium were bombarded by silicon beam, but a variety of observations had been seen as discussed above. Similarly, all other eight targets were collided by oxygen ions, but the results are very different. They can be put in three categories as discussed above also.
\begin{table*}[htbp]
\centering
\caption{Comparison of  the measured $v_F$ (in units of $10^6$ m/s) with different model predictions \cite{Ashcroft76, gall2016electron,montanari2017low} is given. Corresponding number of conduction electron $z$ are also compared. $z$ in the third column represents the value as it comes from the electronic configurations, where as $z^\prime$ in the sixth column and $z^{\prime\prime}$ in the eighth column correspond the average number of conduction electron as obtained from model predicted $v_F$ \cite{gall2016electron,montanari2017low} and the presently measured $v_F$, respectively.} 
\begin{tabular}{|c|l|c|c|l|c|c|c|}
\hline
\textbf{Elm.} & \textbf{Electr. Config.} & ~$z$~ & $v_F$ \cite{montanari2017low,Ashcroft76} & $v_F$ \cite{gall2016electron} & $z'$ & $v_F$ (Pres. Expt.) & $z''$ \\
\hline
Zn  & [Ar] 3d$^{10}$ 4s$^2$                  & 2 & 1.83 \cite{Ashcroft76} & 1.566                         & 1.25 & $1.735 \pm 0.090$  & $1.70^{+0.28}_{-0.25}$ \\
Au  & [Xe] 4f$^{14}$ 5d$^{10}$ 6s$^1$        & 1 & 1.40 \cite{Ashcroft76} & 1.382                          & 0.96 & $1.382 \pm 0.055$  & $0.96^{+0.13}_{-0.11}$\\
Pb  & [Xe] 4f$^{14}$ 5d$^{10}$ 6s$^2$ 6p$^2$ & 4 & 1.83 \cite{Ashcroft76} &                    & 
& $1.830 \pm 0.036$  & $4.00^{+0.29}_{-0.19}$ \\
&&& 1.82 \cite{montanari2017low} &&&&\\

Bi  & [Xe] 4f$^{14}$ 5d$^{10}$ 6s$^2$ 6p$^3$ & 5 & 1.87 \cite{Ashcroft76} &                                          &       & $1.870 \pm 0.044$  & $5.00^{+0.40}_{-0.30}$ \\
\hline

Ge  & [Ar] 3d$^{10}$ 4s$^2$ 4p$^2$           & 4 & 2.01 \cite{Ashcroft76} &                      &     & $2.501 \pm 0.047$  & $7.70^{+0.45}_{-0.45}$ \\
&&&2.00 \cite{montanari2017low} &&&&\\
Ag  & [Kr] 4d$^{10}$ 5s$^1$                  & 1 & 1.39 \cite{Ashcroft76} & 1.440                           & 1.10  & $3.710 \pm 0.090$    & $19.0^{+1.35}_{-1.35}$ \\
Ta  & [Xe] 4f$^{14}$ 5d$^3$ 6s$^2$           & 5 & 2.34 \cite{Ashcroft76} &                                    &   & $3.488 \pm 0.056$  & $16.5^{+0.85}_{-0.70}$ \\
W   & [Xe] 4f$^{14}$ 5d$^4$ 6s$^2$           & 6 & 2.59 \cite{Ashcroft76} & 0.971   & 0.31 & $3.924 \pm 0.064$  & $20.75^{+1.02}_{-0.95}$ \\
\hline
Cu  & [Ar] 3d$^{10}$ 4s$^1$                  & 1 & 1.57 \cite{Ashcroft76} & 1.110                          & 0.35 & $1.110 \pm 0.026$  & $0.35^{+0.03}_{-0.03}$ \\
Th  & [Rn] 6d$^2$ 7s$^2$                     & 2 & 1.40  \cite{Ashcroft76} &                                          &       & $1.402 \pm 0.039$  & $2.00^{+0.15}_{-0.20}$ \\
& & 4 & 1.77 \cite{Ashcroft76}&&&& \\
U   & [Rn] 5f$^3$ 6d$^1$ 7s$^2$              & 2 & 1.64 \cite{Ashcroft76} &                                          &       & $1.632 \pm 0.024$  &$1.95^{+0.10}_{-0.08}$  \\
& & 6 & 2.37 \cite{Ashcroft76}&&&& \\

\hline
\end{tabular}
\label{tab:fermi_velocities}
\end{table*}
\section{Conclusion}
The continuous effort to fully understand the heavy-ion-induced x-ray production cross section data in recent years \cite{chatterjee2021significance, chatterjee2022understanding, kaur2023understanding, kaur2025towards} has led us to lay a good foundation. Working further to achieve higher success, we realized the need for accurate Fermi velocities. A Fermi velocity measurement method was devised using the experimentally obtained heavy-ion-induced x-ray production cross-sectional data. This method is so powerful that it enables us to measure the correct and accurate Fermi velocity for almost every elemental metal in the periodic table. This work not only improves our fundamental understanding better on the complex issue of x-ray emission from heavy-ion collisions but also helps to find a good substitute for copper in further miniaturization of the integrated circuits and to use the heavy-ion-induced x-ray emission is reliable and accurate for elemental analysis.
\section{Data Availability Statement}
Data will be made available on reasonable request.
\appendix
\section{Details of the error calculations}
The total K x-ray production cross section is given by
\begin{equation}
\centering
\sigma^{X}(tot)=\sigma_{DCI-MI}^{X}(tot) + \sigma_{EC}^{X}(tot) \label{fnl-eqn}
\end{equation}
\noindent Here, $X$, DCI-MI and $EC$ stand for x-ray production, direct Coulomb ionization including multiple ionization and electron capture cross section, respectively. 
using the relation between XPCS and IC 
\begin{equation}
\centering
\sigma^X= \omega_k\sigma^I,\label{Eqn:XPCS-IC}
\end{equation}
we can write 
\begin{equation}
  \frac{{\Delta\sigma}^{X}_{(tot)}}{\sigma^{X}_{(tot)}} = \frac{\sigma_{DCI-MI}^{I}}{\sigma^{I}_{(to t)}}\cdot\frac{\Delta\sigma_{DCI-MI}^{I}}{\sigma_{DCI-MI}^{I}}+\frac{\sigma_{EC}^{I}}{\sigma^{I}(tot)}\cdot\frac{\Delta~\sigma_{EC}^{I}}{\sigma_{EC}^{I}}
\end{equation}
$\sigma_{DCI-MI}^{I}$ can be accurately estimated using the ECPSSR theory \cite{kaur2023understanding} and thus the associated error is small and we can assume it within 2\%. The major error comes from $\sigma_{EC}^{I}$ due to uncertainty in the charge state fractions ($F(q)$) inside the target, and it is theoretically estimated from the following electron capture cross section.
\begin{equation}
    \sigma_{EC}^{I} = \frac{2^{p}\pi}{5}\frac{n_1^2}{v_1^2}(\frac{v_{1s^\prime}}{v_{2s}})^5\xi_{ss^\prime}^{10}(\theta_s)\frac{\phi_4(t_{ss^\prime})~\chi_{{ss^\prime}}}{(1+t_{ss^\prime})^3}
\label{Sig-obk-ss'}
\end{equation}
Here, the exponent p is 13, 10 and 8 for K, L and M shell electrons are captured by the projectile ions. The subscripts 1 and 2 refer to the projectile and the target, respectively. The projectile shell $s^\prime$ and the target shell $s$ play an important role in many parameters $\xi_{ss^\prime}$, $\chi_{{ss^\prime}}$, $t_{ss^\prime}$ and $\phi_4(t_{ss^\prime})$ \cite{kaur2023understanding}. Of these, $\chi_{{ss^\prime}}$ is only related to $F(q)$. $v_{1s^\prime}$ is characterized by the quantum number $n_1$ and the orbital velocity of the electron in the projectile is given as $v_{1s^\prime.}(=z_{1}/n_1)$. We can write the error equation as follows:
\begin{equation}
\frac{\Delta\sigma_{EC}^{X}}{\sigma_{EC}^{I}}=\frac{\Delta~\chi_{ss^{\prime}}}{\chi_{ss^{\prime}}} 
\end{equation}
Here, the parameter $\chi_{ss^\prime}$ plays an important role in the ${ss^\prime}$ electron capture processes, and it is given by:
\begin{equation}
\chi_{ss^\prime}=\sum_{q}\frac{F(q)~J_{ss^\prime}(q)}{K_{ss^\prime}(q)}
\label{chi-k}
\end{equation}
$F(q)$ represents the charge-state fraction of the projectile ion within the target charge-state $q$. $J_{ss^\prime}(q)$ denotes the vacancies in the $s^\prime$ shell of the projectile for electron capture from the target $s$-shell at charge state $q$, and $K_{ss^\prime}(q)$ is a constant depending on how many subshells in the projectile $s^\prime$ shell are involved in capture, based on ionization energies. More details can be seen from \citet{kaur2023understanding}.    
\begin{equation}
 \frac{\Delta\chi_{ss^{\prime}}}{\chi_{ss^{\prime}}}=\frac{\sum_{q}\frac{F(q) J_{ss^{\prime}}}{K_{ss^{\prime}}}\frac{\Delta F(q)}{F(q)}}{\sum_{q}\frac{F(q)J_{ss^{\prime}}}{K_{s s^{\prime}}}} 
\end{equation}
$F(q)$ is estimated from the following equation:
\begin{equation}
   F(q)=\frac{1}{\pi}\frac{\frac{\Gamma}{2}}{(q-q_m)^2+(\frac{\Gamma}{2})^2} \:\text{and}\:\sum_q F(q)=1 \label{CSF-eqn} 
\end{equation}
Here, $\Gamma$ and $q_m$ are the width and mean charge state in the charge state distribution. The corresponding error equation is as follows:
\begin{equation}
\frac{\Delta F(q)}{F(q)} = \frac{\Delta \Gamma}{\Gamma}-\frac{4\pi~q_{m}(q-q_{m})}{\Gamma F(q)}\frac{\Delta q_{m}}{q_{m}}-\frac{2\pi~\Gamma}{F(q)} \frac{\Delta \Gamma}{\Gamma},  
\end{equation}
$\Gamma$ is related to $q_m$, $x=\frac{q_{m}}{Z_{1}}$, as follows:
\begin{equation}
\Gamma= C[1-exp(-(x)^\alpha)][1-exp(-(1-x)^\beta)]
\label{Gamma}
\end{equation}
The constants $\alpha$ and $\beta$ are 0.23, 0.32, respectively, and $C = 2.669-0.0098. Z_2+0.058.Z_1+0.00048.Z_1.Z_2$. Now we can write
\begin{equation}
 \frac{\Delta \Gamma}{\Gamma}=-x^{\alpha}e^{-x^{\alpha}}.\begin{matrix}\frac{\Delta x}{x}&+\beta x(1-x)^{\beta-1}-e^{-(1-x)^{\beta}}.\end{matrix}\frac{\Delta x}{x} 
\end{equation}
Note that $\frac{\Delta x}{x}=\frac{\Delta q_{m}}{q_{m}}$ 
$q_m$ is estimated from the Fermi gas model as follows:
\begin{equation}
q_{m}=Z_1 (1-\frac{v_{F}}{v_{1}})     
\end{equation}
Here, $v_F$ is the Fermi velocity. The error in $q_m$ is due to the error in $v_F$ as follows:
\begin{equation}
\frac{\Delta q_{m}}{q_{m}}=\frac{-v_{F}}{v_{1}-v_{f}}\frac{\Delta v_{F}}{v_{F}}.     
\end{equation}
Using the above equations, we can vary $v_F$ to obtain $\sigma_{EC}^I$ or
estimate the measurement uncertainty of $v_F$ from the measured uncertainty of the total XPCS. This procedure can also be applied to the case where L and M x-ray production cross sections are used to measure the Fermi velocity. 

\bibliography{BIB.bib}
\bibliographystyle{apsrev4-1}
\end{document}